\documentclass[onecolumn,usenatbib,amsmath,amssymb,amsbsy]{mn2e}

\usepackage{natbib}
\usepackage{graphicx}

\def\mut{{\tilde{\mu}}}
\def\ap{{\alpha_{+}}}
\def\am{{\alpha_{-}}}
\def\bp{{\beta_{+}}}
\def\bm{{\beta_{-}}}

\newcommand{\be}{\begin{equation}}
\newcommand{\ee}{\end{equation}}
\newcommand{\bear}{\begin{eqnarray}}
\newcommand{\eear}{\end{eqnarray}}

\begin{document} 

\title[A toy model for magnetar oscillations.]
{Elastic or magnetic? A toy model for global magnetar oscillations with 
implications for QPOs during flares.}
  
\author[Glampedakis, Samuelsson \& Andersson]
{Kostas Glampedakis,  Lars Samuelsson and Nils Andersson \\
School of Mathematics, University of
Southampton, Southampton SO17 1BJ, UK}

\maketitle

\begin{abstract}
We use a simple toy-model to discuss global MHD modes of a neutron
star, taking into account the magnetic coupling between the elastic crust
and the fluid core. Our results suggest that the notion of pure torsional 
crust modes is not useful for the coupled system. All modes excite Alfv\'en
waves in the core. However, we also show that the modes that are 
most likely to be
excited by a fractured crust, eg. during a magnetar flare, are such 
that the crust and the core oscillate in concert. For our simple model,
the frequencies of these modes are similar to the ``pure crustal'' 
frequencies. In addition, our model provides a natural explanation for the
presence of lower frequency ($< 30$~Hz) quasi-periodic oscillations seen in 
the December 2004 giant flare of SGR 1806-20.

\end{abstract}


\section{Introduction}
\label{sec:intro}

The recent observational evidence \citep{israel,anna1} of quasi-periodic 
oscillations (QPOs) during 
giant flares in the soft gamma-ray repeaters (SGRs) $1806-20$ and $1900+14 $ may 
herald the beginning
of an exciting new era for neutron star physics (see also \citet{barat} for older evidence of a
QPO in the flare of SGR 0526-66). These extremely violent events 
can be understood within the  
``magnetar'' model, introduced by Duncan and Thompson more than a decade ago 
\citep{DT,TD}.
Magnetars are neutron stars endowed with powerful magnetic fields ($\sim 10^{14-
15}$ G) 
masquerading as both SGRs and anomalous X-ray pulsars. The  giant flare events 
are thought to be 
due to magnetic instabilities leading to fracturing of the crust.
The observed QPOs are then associated with shear oscillations of the magnetar's 
crust \citep{duncan}.
This interpretation makes sense since the 
fundamental toroidal crustal modes have frequencies ($\sim 30-100$~Hz) that could match the 
observations
\citep{vanhorn}. 
The properties of crustal modes have been investigated by several authors, see 
for example the work by 
\citet{hansen, mcdermott, strohmayer, messios, piro}.   

In previous studies the basic fact that the 
magnetic field strongly couples the crust to the fluid core has been ignored. 
This key point, which is well known from discussions of pulsar 
glitch relaxation \citep{easson,abney}, was recently emphasised by 
\citet{levin}.
A rough estimate for the crust-core coupling timescale is provided by 
the Alfv\'en crossing time, 
\be
t_{\rm A} = 2R/v_{\rm A} \sim 70\, B_{15}^{-1}\,\rho_{14}^{1/2}\, R_6 ~\mbox{ms}
\ee
where the magnetic field, density and radius are normalised according to 
$B=10^{15} B_{15}$ G, 
$\rho = 10^{14} \rho_{14}~\mbox{g}/\mbox{cm}^{3} $ and $R=10^6 R_{6} $ cm, 
respectively. 
We have also introduced the Alfv\'en velocity $v_{\rm A}^2 \equiv B^2/4\pi\rho 
$. 
The coupling timescale is comparable to the period of the fundamental crustal 
mode
\citep{mcdermott,hansen},  
\be
P_{\ell}^{0} \approx  60\, R_6\,[\ell(\ell+1)]^{-1/2} ~\mbox{ms}
\label{modep}\ee
This means that, for parameters relevant to the crust-core interface of a 
magnetar, 
an efficient coupling to the fluid core is already established within a single 
oscillation 
of a ``crustal'' mode. This  has crucial 
implications when one attempts to calculate the properties of the modes. In 
essence, the notion of modes
confined to the crust is no longer useful and one is  forced to consider {\it 
global} 
oscillations of the coupled crust-core system. 

In this Letter we demonstrate that a magnetically coupled 
crust-core system admits oscillations with frequencies that  
matches the observational data. Clearly, any initial disturbance in the crust 
region (say,  following
a starquake induced by a magnetic field eruption) will generically shake the 
field lines and 
launch Alfv\'en waves into the core. In contrast to \citet{levin}, we do not 
think of this process as a damping mechanism.
The reason for this is that the Alfv\'en waves in the core will be capable of 
reaching the crust at 
the opposite side of the star and be reflected back. By the time that a given 
wave-packet 
is eventually attenuated by viscosity it would have transversed the core some 
$\sim 10^7$ times. 
Hence, we believe that the correct approach to the problem is to consider the 
coupled crust-core system and 
compute global oscillation modes.


\section{A simple toy model}

We consider a plane-parallel ``star'' where the fluid core is sandwiched by two
slabs of ``crust'' (this model is similar to the one employed by  \citet{piro}). 
The $z$ coordinate runs 
from $z=+R $ (the surface) to $z=0 $ (the core's centre) and ends up back at the 
surface, at $z=-R $. 
The crust-core interface is located at $z=\pm R_c $. The crust is elastic, with
a uniform shear modulus $\mu$. Furthermore, uniform density, incompressibility 
and 
ideal MHD conditions are assumed everywhere. In the unperturbed configuration 
the magnetic field is 
${\bf B} = B_\circ {\bf \hat{z}} $. 

The Euler equation in the crustal region is (assuming a harmonic time dependence 
${\bf \xi} \sim \exp(i\sigma t)$),
\be
-\sigma^2 \xi_i = \frac{1}{\rho} \frac{\partial \sigma_{ik}}{\partial x^k} 
+ \frac{1}{4\pi\rho} \left [\, (\nabla \times {\bf b} ) \times {\bf B_\circ} 
\right ]_i
\ , \qquad \mbox{ where } \quad 
\sigma_{ij} = \mu \left [ \frac{\partial \xi^i}{\partial x^j}
+ \frac{\partial \xi^j}{\partial x^i} \right ] 
\label{euler1}
\ee
$\sigma_{ik}$ and $ {\bf b}$  are the perturbed shear tensor and magnetic field,
respectively. Since the fluid is incompressible and all  quantities vary only 
with respect
to $z$ we have,
\be
{\bf b} = B_\circ [\,{\bf \hat{x}} \partial_z \xi^x + {\bf \hat{y}} \partial_z 
\xi^y\, ] 
\ee

Following \citet{piro}, we  mimic the geometrical factors of the true spherical 
problem  by making the identification $(\partial_x^2 + \partial_y^2){\bf \xi} \to -\ell(\ell+1){\bf 
\xi}/R^2 $. To introduce  $\ell$ in this way is, of course, artificial. In the real 
spherical problem, the $\ell$-multipoles follow from separation of variables. Here 
they are simply introduced to impose the ``correct scaling'' for the crust mode frequencies.
Eqn.~(\ref{euler1}) now becomes
\be
\left [\,\mut + v_A^2 \, \right ] \partial_z^2 \xi_i + \left [\, \sigma^2 -
\mut\frac{\ell(\ell+1)}{R^2} \,
\right ] \xi_i = 0, \quad i = x,y
\label{euler2}
\ee
where $\mut = \mu/\rho $. 
Clearly, identical equations describe $\xi^x $ and $\xi^y $, therefore we
shall only solve for the former. In addition, as our model has reflection 
symmetry with
respect to the ``centre'' $z=0 $. Hence it is in principle sufficient to solve 
only for $z > 0$. 

The boundary conditions are formulated in terms of the traction components
\citep{strohmayer,carroll,piro}. For our combined fluid/magnetic field system 
the traction takes
the following form in the crust region:
\be
T^i = \rho [ \mut + v_A^2 ] \partial_z \xi^i, \quad i = x,y \qquad \mbox{and} 
\qquad T^z = 0 
\label{tract1}
\ee
The corresponding expressions for the core region are obtained 
by setting $\mut = 0 $. 
Another important element is the perturbed electric field,
\be
{\bf e} = -\frac{i\sigma}{c} {\bf \xi} \times {\bf B_\circ} =
\frac{i\sigma B_\circ}{c} [ -{\bf \hat{x}} \xi^y +   {\bf \hat{y}} \xi^x ]
\ee
The appropriate MHD boundary conditions are: (i) continuity of the tractions at 
the 
crust/core interface, (ii) vanishing tractions at the surface and (iii) 
continuity
of the (normal) transverse components of the (magnetic) electric field at the 
interface. 
These translate into,
\be
\xi^x (R_c^{+}) = \xi^x (R_c^{-}), \qquad
\partial_z \xi^x (R) = 0 \qquad \mbox{and} \qquad
\partial_z \xi^x (R_c^{-}) = \left [  1+ \mut/v_A^2 \right ]\,
\partial_z \xi^x (R_c^{+}) 
\label{bc0}
\ee 
These constraints imply that the displacement 
${\bf \xi}$ is a continuous function at the crust-core interface. 
Note also that, in solving the problem one has two options. 
One can either solve the full problem and impose conditions analogous to (\ref{bc0}) 
 for negative $z$. Alternatively, one can use the symmetry of the problem and impose a 
suitable condition on the solution at the origin (the eigenfunction should be
either an odd or an even function). 


We begin by considering the  special case where the core is
decoupled from the crust. As we will see later, this case is artificial 
in the presence of a magnetic field. 
However, the magnetic coupling between the crust and the fluid core  
can be deactivated by altering the boundary conditions at the bottom of the 
crust \citep{carroll,piro,messios}.

Solving the Euler equation in the crust provides the displacement,
\be
\xi^x = C_1\,e^{\ap z} + C_2\,e^{\am z}, 
\qquad \mbox{where} \qquad 
\alpha_{\pm} =\pm \frac{1}{R} \left [ \frac{\ell(\ell+1)  -\tilde{\sigma}^2}{1 +
v_A^2/\mut} \right ]^{1/2} \quad \mbox{and} \quad 
\tilde{\sigma} \equiv \sigma \frac{R}{\mut^{1/2}}
\label{crust_xi}
\ee
The effective decoupling between the core and the crust can be achieved by 
requiring 
vanishing tractions at the interface. We stress the fact that this is {\it not} the
correct interface condition in the magnetic problem. Disregarding this we get,
\be
C_2= C_1\, e^{2R_c \ap}
\label{bc1}
\ee
The requirement of vanishing tractions at the surface gives a similar
relation
\be
C_2 = C_1\, e^{2R \ap}
\label{bc2}
\ee
These two expressions are compatible provided that
\be
e^{2\Delta\ap} = 1 \quad \Rightarrow\quad  \sigma_n = \frac{\mut^{1/2}}{R}\,
\left [ \, \ell(\ell+1) + \left (\frac{n\pi R}{\Delta}\right )^2 
\left \{ 1 + \frac{v_A^2}{\mut} \right \}
\right ]^{1/2} \quad \mbox{with} \quad n = 0,1,2,3,... 
\ee
where $\Delta = R -R_c $ is the crust thickness. This is the prediction of our 
toy model for the ``toroidal'' crust mode frequencies. It is in good agreement with 
more rigorous results \citep{mcdermott,hansen,piro}. However, the model is inaccurate
for $\ell=1$ since a nodeless eigenfunction is allowed, a feature not
found in the rigorous spherical problem. Given the artificial introduction of $\ell$ in the model, 
this is not surprising. For $\ell \geq 2 $ (and $\ell=1,~ n \geq 1 $) the model predicts the correct 
number of nodes for the eigenfunctions. It is useful to note that the fundamental ($n=0 $) and 
first overtone ($n=1 $) quadrupole modes take the respective values 
$\tilde{\sigma}_{0} = 2.45 $ and $\tilde{\sigma}_{1} = 31.5 $ (for the latter
we have assumed $\Delta=0.1 R$ and $v_{\rm A}=0 $).

In the case of the  full crust-core model we solve the Euler equations in the 
two regions. This (again) leads
to  a solution of the form (\ref{crust_xi}) in the crust, while in the core we 
find  
\be
\xi^x = D_1\,e^{i \bp z} + D_2\,e^{i \bm z}, 
\qquad \mbox{where} \qquad 
\beta_{\pm} =\pm\, \sigma/v_{\rm A} 
\label{core_xi}
\ee 
Once more, the surface boundary conditions enforce (\ref{bc2}). On the other
hand, eqn.~(\ref{bc1}) is no longer valid. Instead, the appropriate interface 
conditions
result in a $2\times 2 $ homogeneous system for the coefficients $D_1$ and 
$D_2$.
The requirement that the relevant determinant must vanish provides the following
condition
\be
e^{-2iR_c \bp}\, \left [ \ap  \left (1 + \frac{\mut}{v_A^2} \right )
\sinh (\Delta \ap) -i\bp \cosh(\Delta \ap)  \right ] ^2
- e^{2iR_c \bp}\, \left [ \ap  \left (1 + \frac{\mut}{v_A^2} \right )
\sinh (\Delta \ap) +i\bp \cosh(\Delta \ap)  \right ] ^2 = 0
\label{spectrum}  
\ee
which determines the complete spectrum of global oscillations.
For the amplitudes we find,
\be
D_1 =-C_1 \frac{i\ap}{\bp}\,e^{R_c \ap }\left [1+ \frac{\mut}{v_A^2}  \right ]
\left [\,\frac{1-e^{2\Delta \ap}}{e^{i\bp R_c} \mp e^{-i\bp R_c}} \, \right 
] \qquad \mbox{ and } \qquad 
 D_2 = \pm D_1 
\ee
(where the upper/lower signs correspond to the same solution). 
Up to normalisation, this completely specifies the solution. It is clear that the 
modes are either odd or even functions with respect to the origin.
It is also worth pointing out that, 
since the limit $v_{\rm A}^2 \to 0 $ is singular, some care is required
in analysing how the pure crust modes emerge as the magnetic field vanishes.

 
\section{Results: Global MHD modes}

Despite being very simple, the toy-model provides a set of interesting results.
First note that we can obtain analytic solutions to (\ref{spectrum}) 
for a weak magnetic field. When $v_A^2 \ll \mut $ acceptable mode solutions 
coincide with 
the {\it Alfv\'en frequencies}:
\be
\sigma_A = (k\pi/2 R_c) \,v_{\rm A}, \quad k = 0,1,2,3,... \quad
\label{freqA}
\ee
Remarkably, these frequencies also satisfy (\ref{spectrum})
when $e^{2\Delta\ap} = 1 $, that is, when the frequency coincides with
a crustal frequency $\sigma_n $. This triple intersection is naturally interpreted as a 
resonance between the crust and the core.
 
\begin{figure}
\centerline{\includegraphics[height=5.5cm,clip] {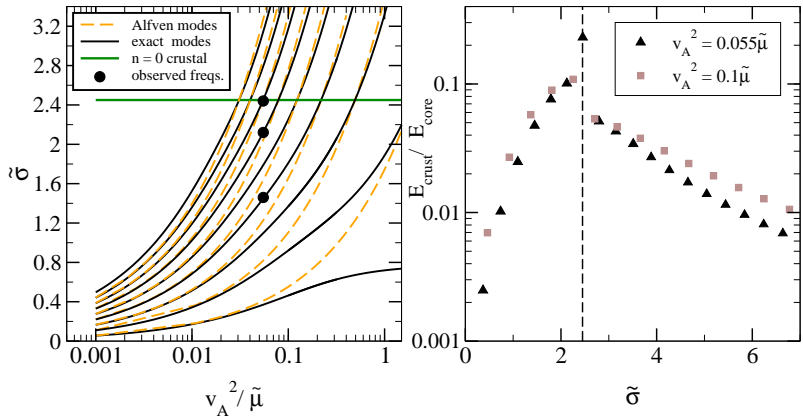}}
\caption{{\it Left panel}: A part of the global mode spectrum for the toy model 
of a magnetar with crust-core coupling, 
see text for details. The data corresponds to  $\ell=2 $ and $\Delta = 0.1 R $. 
Note the triple intersection between the exact mode solution from 
eqn.~(\ref{spectrum}), 
the Alfv\'en modes from eqn.~(\ref{freqA}), and the fundamental crustal 
frequency 
$\tilde{\sigma}_{0} = 2.45$. The black dots correspond to 
observed QPOs frequencies for SGR 1806-20,
see text for discussion. {\it Right panel}: Assessing the excitation of various 
modes by means
of the ratio between the mode energy in the crust $E_{\rm crust}$ and that in 
the core, $E_{\rm core} $. The dashed line labels the crustal 
frequency $\tilde{\sigma}_{0}$.}
\label{fig1}
\end{figure}

We have determined the exact spectrum by solving (\ref{spectrum}) numerically. 
A part of the  spectrum for $\ell=2$ and $\Delta=0.1 R $ is shown in the left 
panel of 
Fig.~\ref{fig1}, together with the Alfv\'en frequencies (\ref{freqA}), as a 
function of the ratio
\be
v_{\rm A}^2/\mut \approx 0.04\, B_{15}^2\, \rho_{14}^{-1}
\label{ratio}
\ee  
We have assumed the value 
$\mut \approx 2\times 10^{16} \mbox{cm}^2/\mbox{s}^2 $ 
\citep{douchin}. 
Figure~\ref{fig1} shows that, for a given value of the ratio (\ref{ratio}), the 
spectrum consists of a semi-infinite family of modes (we only show the first few). The
separation between consecutive modes increases as the ratio attains higher 
values, i.e. if we increase the magnetic field while keeping the other parameters fixed. 
The most important feature to note in the spectrum is that there are always mode-frequencies 
comparable to the crustal frequencies $\sigma_n$. Fig.~\ref{fig1} depicts the spectrum in the 
vicinity of $\sigma_{0} $ but a similar picture arises for {\it all} frequencies
$\sigma_n$ of a given $\ell$. As is apparent in the figure, for certain discrete values of the 
ratio $v_{\rm A}^2/\mut$ the mode frequency {\it exactly} coincides with a crustal frequency 
$\sigma_n$ as well as one of the Alfv\'en frequencies (\ref{freqA}). For any other realistic value 
of $v_{\rm A}^2/\mut$ there is always a mode with frequency very close (with at most a few 
percent deviation) to each ``crustal'' frequency $\sigma_n$.

\begin{figure}
\centerline{\includegraphics[height=4.3cm,clip] {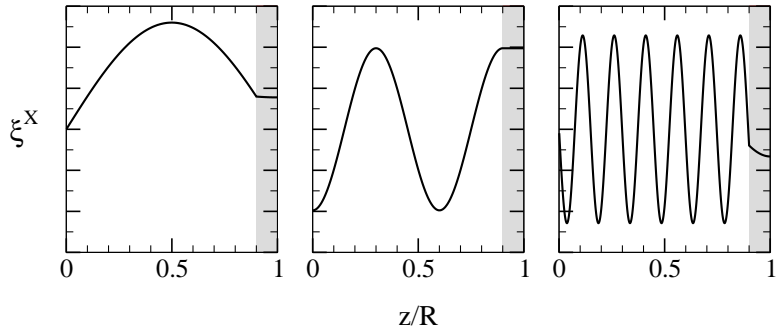}}
\caption{Fluid displacement $\xi^x $ for $\ell=2,~\Delta=0.1 R$ and 
$v_{\rm A}^2 = 0.055\,\mut $. The modes illustrated are, from left to right: 
$\tilde{\sigma} = 0.74, 2.45, 9.87$. In the middle panel
note the comparable oscillation amplitudes in the crust and the core 
for the mode with $\sigma \approx \sigma_{0} = \sigma_{\rm A}$.}
\label{fig2}
\end{figure}

The coupled crust-core system obviously 
has a much richer set of oscillation modes than the decoupled elastic crust. Yet 
one can
interpret most of the observed QPOs as pure crust modes for various values of 
$\ell$, 
see the discussion of \citet{anna1}.
At first sight our results may seem at variance with this interpretation. 
After all, our system  has a number of additional modes. If they are not 
observed we need
to explain why. To do this, we take the standard model for the giant flares at 
face value. 
The magnetic field erupts and induces a starquake in the crust. On the grounds 
of physical intuition we
would
expect that the initial perturbation  in the crust will predominantly excite 
those global modes which 
communicate the least amount of energy to the core. To test this idea, we 
calculate the
total energy (kinetic + magnetic) associated with each mode and consider the 
ratio  $E_{\rm crust}/E_{\rm core} $.
As can be seen from the right panel of Fig.~\ref{fig1}, the results corroborate 
our intuition. The mode which {\it maximises} the energy ratio, i.e. should be 
easier to 
excite from an initial crust motion,  is the one nearest to the resonant frequency 
$\sigma_{\rm A}= \sigma_{0}$ (analogous behaviour to that depicted in the right panel of
Fig.~\ref{fig1} is found for modes in the vicinity of all higher $\sigma_n $ frequencies, 
for any $\ell$). This would explain why the other modes of the system are more difficult to 
excite, they are predominantly core-Alfv\'en modes which would be energetically more 
expensive to excite via the proposed mechanism. 
Similar conclusions can be drawn from the displacement $\xi^x $, displayed in
Fig.~\ref{fig2}, for a fixed value of $v_{\rm A}^2/\mut $, as we move upwards in the spectrum. 
The oscillation amplitude is generally far greater in the core, 
but the modes located in the vicinity of the crust frequencies $\sigma_n$ are exceptions. In 
those cases the crust oscillates with a comparable amplitude. However, these modes cannot be 
considered as ``crustal''. The eigenfunctions clearly show that they represent global oscillations. 

The conclusion of this discussion is that once the crust is shaken by a starquake the core-crust system 
will naturally choose to vibrate in those global modes which have frequencies similar to the frequencies of the 
toroidal modes of the uncoupled crust, with identical values of $\ell$ (and number of nodes in the crust). 
Hence, the model provides an explanation for all observed QPO frequencies, both intermediate 
($\sim 30-155~$ Hz, different values of $\ell$) and 
high frequency ($\sim 625~$ Hz, overtone with one node in the crust). 
Our model differs from previous ones only in that the modes are global, not 
localised to the crust, and hence have a significant amplitude in the core.   

A key feature  of Figure~\ref{fig1} is the existence of modes with frequencies
{\it below} the fundamental crustal frequency $\sigma_{0} $. 
This is interesting since the presence of low frequency QPOs 
has  been confirmed in the data of both the RXTE and RHESSI satellites 
\citep{israel,anna2}: 
$18,26~$ and $30$ Hz QPOs for the December 2004 giant flare in SGR 1806-20. It 
is natural to 
identify the 30~Hz QPO with the  
fundamental crustal frequency $\sigma_{0}$ \citep{israel}, since the higher 
frequency QPOs 
then fit the predictions from Eq.~(\ref{modep}) quite well. However,  
 we then find ourselves left with a puzzle: what is the origin 
of the remaining two frequencies? There are certainly no toroidal crustal modes 
with  frequency 
below $\sigma_{0}$. Our model offers a natural explanation for these 
low-frequency modes. 

Although our model is not designed to provide us with quantitatively accurate 
predictions,
we can still attempt to match the observed data for the low frequency QPOs of 
SGR 1806-20.
To do this we first identify $\tilde{\sigma}_{0} = 2.45 $ with the $30 $ Hz 
frequency 
(presumably, this identification can be made rigorous in a calculation for a 
realistic neutron 
star model). Then the frequencies $18$ and $26$ Hz correspond to modes with 
$\tilde{\sigma}=1.46 $
and $2.12 $, respectively. Some navigation in Fig.~\ref{fig1} leads us to the 
value $v_{\rm A}^2 = 0.055\mut $
which has all three desired modes (indicated by filled circles in the figure). 
This value is reasonably consistent with $B_{15} \approx 1 $ 
assuming $\rho_{14} \approx 1 $ (note that the magnetic field estimated from the  
spin-down
rate for SGR 1806-20 is $7.8\times10^{14}$~G \citep{woods}.  Moreover, the energy 
argument indicates that it 
is reasonable to expect these QPOs to be excited, cf. Fig.~\ref{fig1}. Finally, 
we have an additional mode at $\approx 22$~Hz. In a sense, this 
could be considered a testable prediction.
It would certainly be rewarding if a QPO  with this intermediate frequency 
were to be found in the data!

\section{Concluding remarks}

We have used  a simple plane-parallel toy-model to discuss global MHD modes of a 
neutron
star. The model, which takes into account the magnetic coupling between the 
elastic crust
and the fluid core, provides results that are highly suggestive. 
The system is characterised by a rich spectrum of
global modes ``living'' both in the crust and core, with a mixed magneto-elastic 
identity. Among myriads of 
available modes (for a given multipole  $\ell $ and ratio of Alfv\'en velocity 
to shear velocity) the system
will naturally excite the mode(s) which transfer as little energy as possible to 
the core, 
thus minimising the overall energy budget.
We demonstrated that these favoured global oscillation modes
have frequencies that are similar to the ``pure crustal'' 
frequencies from the non-magnetic problem. Our model provides support for the 
asteroseismology
interpretation for the observed magnetar QPOs.  
In addition, it provides a natural explanation for the
presence of lower frequency ($< 30$~Hz) QPOs seen in the December 2004 giant
flare of SGR 1806-20.

Even though our model is based on drastic simplifications, the basic physics 
predicted should survive for more realistic neutron star models. Obviously, one must expect 
corrections (and extra technical complications!) once the real problem is considered. A more complex  
magnetic field configuration will lead to intricate coupling between different multipoles, 
non-uniform density/crust elasticity and core stratification leading to buoyancy effects are 
likely to affect the mode structure, and the expected superconductivity will affect that nature of the Alfven waves. 
Nevertheless, global MHD modes due to the inescapable crust-core coupling are still likely 
to emerge, leading to intricate resonance phenomena. 
   
%

\section*{Acknowledgements}

We thank Anna Watts for helpful discussions.  This work was supported
by PPARC through grant number PPA/G/S/2002/00038 (KG and NA) and by the EU Marie
Curie contract MEIF-CT-2005-009366 (LS).  NA also acknowledges support from
PPARC via Senior Research Fellowship no PP/C505791/1.




\begin{thebibliography}{}

\bibitem[\protect\citeauthoryear{Abney et al}{{Abney et al}}{1996}]{abney}
Abney M., Epstein R.I., and Olinto A., 1996, ApJ, 466, L91  

\bibitem[\protect\citeauthoryear{Barat et al}{{Barat et al}}{1983}]{barat}
Barat C. et al, 1983, A \& A, 126, 400 

\bibitem[\protect\citeauthoryear{Carroll et al}{{Carroll et al}}{1986}]{carroll}
Carroll B.W. et al, 1986, ApJ, 305, 767 

\bibitem[\protect\citeauthoryear{Douchin \& Haensel}{{Douchin \& 
Haensel}}{2001}]{douchin} 
Douchin F. and Haensel P., 2001,  A \& A,   380,  151 

\bibitem[\protect\citeauthoryear{Duncan \& Thompson}{{Duncan \& 
Thompson}}{1992}]{DT}
Duncan R.C. and Thompson C., 1992, ApJ, 392, L9

\bibitem[\protect\citeauthoryear{Duncan}{{Duncan}}{1998}]{duncan}
Duncan R.C., 1998,  ApJ,  498, L45

\bibitem[\protect\citeauthoryear{Easson}{{Easson}}{1979}]{easson}
Easson I., 1979, ApJ, 228, 257 


\bibitem [\protect\citeauthoryear{Hansen \& Cioffi}{{Hansen \& Cioffi}}{1980}] 
{hansen}
Hansen C.J. and Cioffi D.F., 1980,  ApJ, 238, 740

\bibitem[\protect\citeauthoryear{Israel et al}{{Israel et al}}{2005}]{israel}
Israel G.L. et al,  2005, ApJ,  628, L53

\bibitem[\protect\citeauthoryear{Levin}{{Levin}}{2006}]{levin}
Levin Y., 2006, MNRAS, 368, L35  

\bibitem [\protect\citeauthoryear{McDermott et al}{{McDermott et 
al}}{1988}]{mcdermott}
McDermott P.N., Van Horn H.M., and Hansen C.J., 1988,  ApJ, 325, 725 

\bibitem[\protect\citeauthoryear{Messios et al}{{Messios et al}}{2001}]{messios}
Messios N., Papadopoulos D.M., and Stergioulas N., 2001, MNRAS, 328, 1161

\bibitem[\protect\citeauthoryear{Piro}{{Piro}}{2005}]{piro}
Piro A.L., 2005, ApJ, 634, L153 


\bibitem[\protect\citeauthoryear{Strohmayer}{{Strohmayer}}{1991}]{strohmayer}
Strohmayer T.E., 1991, ApJ,  372, 573 

\bibitem[\protect\citeauthoryear{Strohmayer \& Watts}{{Strohmayer \& 
Watts}}{2005}]{anna1}
Strohmayer T.E., and Watts A.L., 2005, ApJ,  632, L111

\bibitem[\protect\citeauthoryear{Thomson \& Duncan}{{ Thomson \& 
Duncan}}{1995}]{TD}
Thomson C., and Duncan R.C., 1995, MNRAS, 275, 255 

\bibitem[\protect\citeauthoryear{Van Horn}{{Van Horn}}{1980}]{vanhorn}
Van Horn H.M., 1980, ApJ, 236, 899 

\bibitem[\protect\citeauthoryear{Watts \& Strohmayer}{{Watts \& 
Strohmayer}}{2006}]{anna2}
Watts A.L., and Strohmayer T.E., 2006, ApJ, 637, L117 

\bibitem[\protect\citeauthoryear{Woods \& Thompson}{{Woods \& Thompson}}{2004}]{woods}
Woods P.M., and Thompson C., 2004, in ``Compact Stellar X-ray Sources'' eds. W.H.G. Lewin and
M. Van der Klis (preprint astro-ph/0406133)
 




\end{thebibliography}
\end{document}